\documentclass[a4paper]{article}

\usepackage{INTERSPEECH2021}
\usepackage{cite}
\usepackage{amsmath,amssymb,amsfonts}
\usepackage{algorithmic}
\usepackage{graphicx}
\usepackage{textcomp}
\usepackage{xcolor}
\usepackage{lipsum}
\usepackage{multirow}
\usepackage{balance}
\usepackage[inline]{enumitem}
\usepackage{url}
\usepackage{verbatim}
\usepackage{pbox}

\title{Towards Interpretable and Transferable Speech Emotion Recognition: Latent Representation Based Analysis of Features, Methods and Corpora}
\name{Sneha Das$^{1,2}$, Nicole Nadine Lønfeldt$^2$, Anne Katrine Pagsberg$^{2, 3}$, Line H. Clemmensen$^1$}
\address{
  $^1$Department of Applied Mathematics and Computer Science, Technical University of Denmark\\
  $^2$Child and Adolescent Mental Health Center, Copenhagen University Hospital, Capital Region\\
  $^3$Faculty of Health, Department of Clinical Medicine, Copenhagen University}
\email{sned@dtu.dk, nicole.nadine.loenfeldt@regionh.dk, Anne.Katrine.Pagsberg@regionh.dk, lkhc@dtu.dk}

\begin{document}
\maketitle

\begin{abstract}
In recent years, speech emotion recognition~(SER) has been used in wide ranging applications, from healthcare to the commercial sector. In addition to signal processing approaches, methods for SER now also use deep learning techniques. However, generalizing over languages, corpora and recording conditions is still an open challenge in the field. Furthermore, due to the black-box nature of deep learning algorithms, a newer challenge is the lack of interpretation and transparency in the models and the decision making process. This is critical when the SER systems are deployed in applications that influence human lives. In this work we address this gap by providing an in-depth analysis of the decision making process of the proposed SER system. Towards that end, we present low-complexity SER based on undercomplete- and denoising- autoencoders that achieve an average classification accuracy of over 55\% for four-class emotion classification. Following this, we investigate the clustering of emotions in the latent space to understand the influence of the corpora on the model behavior and to obtain a physical interpretation of the latent embedding. Lastly, we explore the role of each input feature towards the performance of the SER.\footnote{For reproducibility the code will be made available at this link by the time of publication: \url{https://bit.ly/3fDWSbq}}
\end{abstract}

\noindent\textbf{Index Terms}: Speech emotion recognition, Autoencoder, Interpretability, Knowledge transfer, Transparency.

\section{Introduction}
Speech emotion recognition~(SER) refers to a group of algorithms that deduce the emotional state of an individual from their speech utterances. SER combined with affect recognition, which uses other modalities like vision and physiological signals are deployed in a wide range of applications. For instance, in the detection and intervention of disorders in healthcare, monitoring the attentiveness of students in schools, risk assessment within the criminal justice system, and for commercial applications, like detecting customer satisfaction in call-centers and by employment agencies to find suitable candidates~\cite{jamaPaed1, kairos1}. 

State-of-the-art SER techniques have evolved from the more conventional signal processing and machine learning based methods to deep neural network based solutions~\cite{akccay2020speech}. Classical methods were based on hidden Markov models~(HMM), Gaussian mixture models~(GMM), support vector machines~(SVM) and decision trees. Contributions based on HMM employed energy and pitch features~\cite{hmm1}, and log-frequency power coefficients from the spectrum~\cite{hmm2} and showed high classification accuracy. Spectral, prosodic and energy features in tandem with a GMM and SVM were used to recognize emotions from Basque and Chinese datasets~\cite{luengo2005automatic, GMM_supervector}, while pitch based input features were used to obtain a GMM and tested on more heterogeneous speech corpora~\cite{busso_GMM}. SVM is another popular method, either used as the primary classification tool, or in coordination with other techniques to predict the affect classes~\cite{friedman2001elements}. 
Recently, more advanced methods that use deep learning have been used for SER. Long short-term memory~(LSTM), bidirectional LSTM and recurrent neural networks~(RNN) were used to predict the quadrant in the dimensional emotional model~\cite{lstm_rnn_blstm}. Following this, denoising autoencoders~(DAEs)were used to learn a lower dimensional latent representation for the emotions that were then employed at various levels to classify speech into emotional categories~\cite{deng2013sparse, xia2013using, ae_domainAdapt}. Convolution neural networks, having had immense success in computer vision, are a common architecture choice for neural network based SERs. 
\begin{figure}[!t]
\centering
\includegraphics[width=0.9\columnwidth]{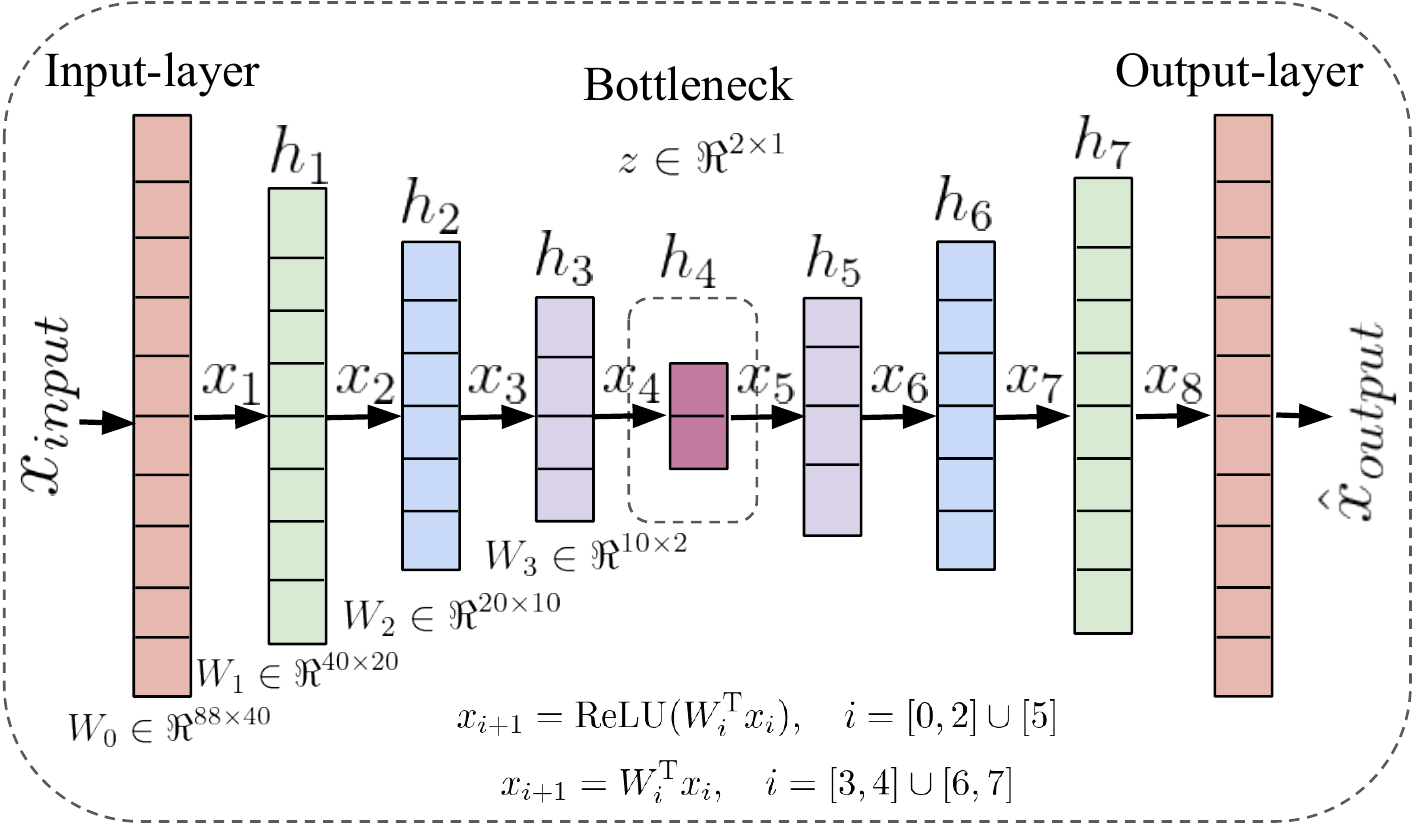}
\caption{Illustration of the autoencoder architecture.}
\label{fig:archi}
\end{figure}

Despite the long history of research contributions in the domain, state-of-the-art methods often struggle to generalize, across corpora with different languages, recording conditions, cultures. Shortage of annotated data in lesser-spoken languages further amplifies issue. Furthermore, many deep learning based SERs are highly complex black-boxes whereby the models are not interpretable and they lack the transparency in their decision making process. This is a crucial aspect, especially when the technology is deployed in applications with consequences on people's lives and access to resources~\cite{AINow2019}.

Unsupervised learning is a promising approach to address label shortage in developing SERs. In addition, autoencoders~(AEs), an unsupervised learning technique, and latent representation studies are useful tools in making models more interpretable. This can also lead to better knowledge transfer between data sets. However, despite the use of AEs for SER in existing literature, few methods provide insights beyond classification accuracy. In this work: \begin{enumerate*}\item we present a low-complexity undercomplete AE~(UAE) and DAE for SER that achieves a performance similar to existing methods, \item we show that the clusters in the latent space implicitly model the activation variable from the dimensional emotional model, \item we study the robustness of the methods by investigating the differences that occur in the latent representations when the underlying data conditions are modified. In other words, how the differences in the language of the corpus impacts the latent space, hence the performance. \item We introduce interpretation in the system by analysing the feature attributions towards emotion clustering. The DeepLIFT algorithm~\cite{shrikumar2017learning} is used to gain insights into the feature subsets that contribute most towards the discrimination of the emotion classes. \end{enumerate*}

\noindent
{\bf Related Work:}
Unsupervised learning techniques for SER have received a lot of interest recently, and AEs are commonly used methods for unsupervised learning. DAE was one of the earliest deep learning based unsupervised learning techniques for SER~\cite{xia2013using}. This was followed by the use of sparse AE for feature transfer~\cite{deng2013sparse} and for SER on spontaneous data set~\cite{dissanayake2020speech}. Furthermore, end-to-end representation learning for affect recognition from speech was proposed and showed performance comparable to existing methods~\cite{ghosh2016representation, ghosh2015learning}. In recent years, techniques like variational and adversarial AEs and adversarial variational Bayes have been exploited to learn the latent representations of speech emotions with input features ranging from the raw signals to hand crafted features~\cite{latif2018variational, parthasarathy2019improving, eskimez2018unsupervised, neumann2019improving}.

\begin{figure}[!t]
\centering
\includegraphics[width=0.9\columnwidth]{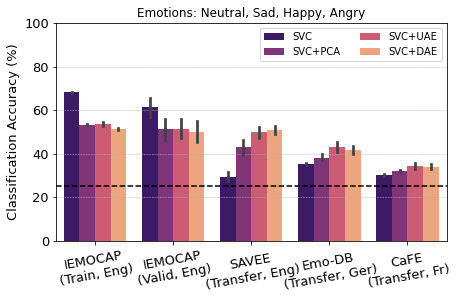}
\caption{Mean and 95\% confidence intervals of the 4-category classification accuracy from 10-fold cross-validation and the dashed line represents random classification level.}
\label{fig:acc_emo4}
\end{figure}

\begin{figure*}[!t]
\centering
\includegraphics[width=0.99\textwidth]{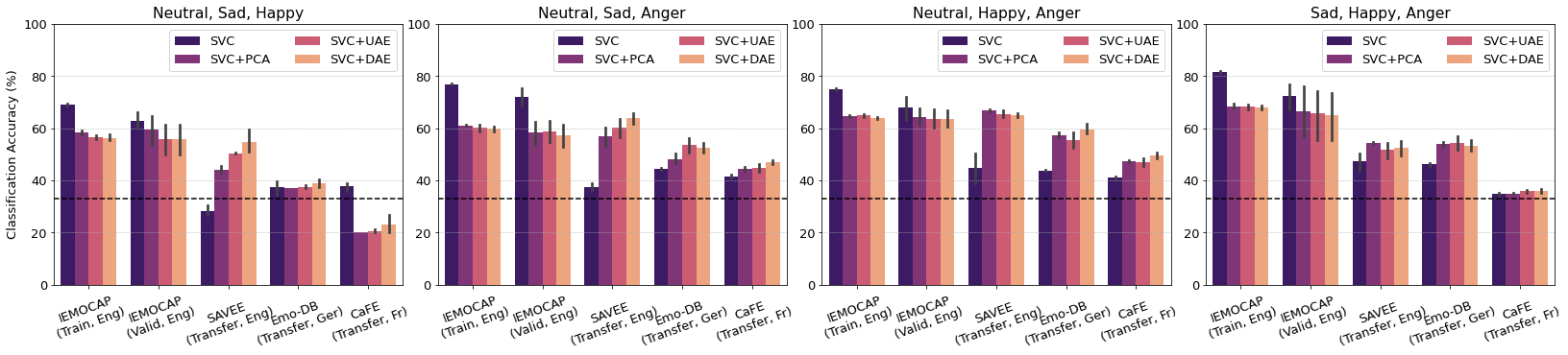}\\
\includegraphics[width=0.99\textwidth]{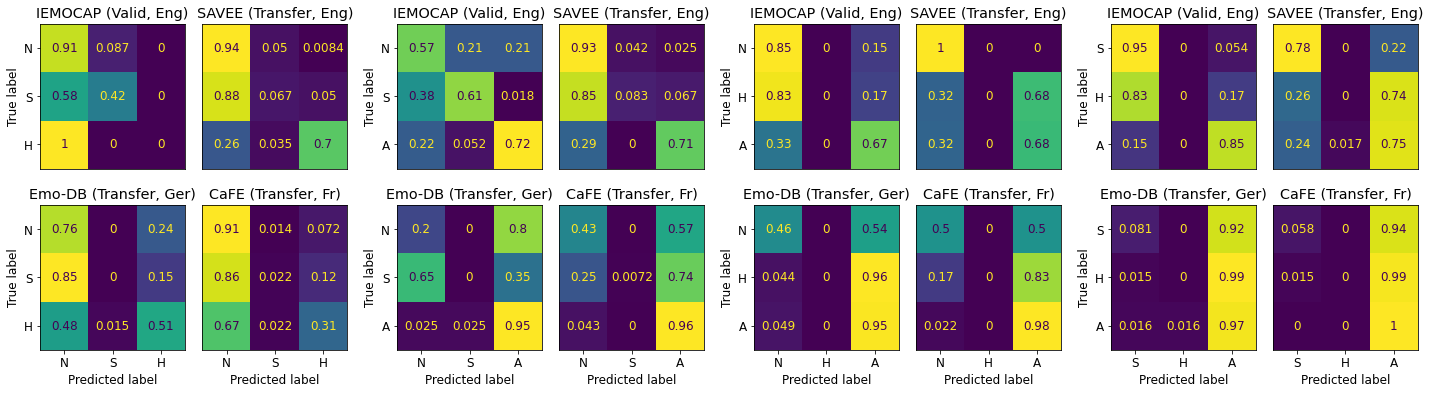}
\caption{Mean and 95\% confidence intervals of the classification accuracy from 10-fold cross-validation over train, valid and transfer data sets. The confusion matrix of the denoising autoencoder for an arbitrary fold. The dashed line represents random classification.}
\label{fig:acc_emo3}
\end{figure*}
\section{Methods and Experimental Setup}
In this section, we provide an overview of the features, the data sets, the architecture of the proposed algorithms and briefly discuss the reference algorithms that are employed in this work. 

\noindent
{\bf Data sets:} 
We use the IEMOCAP data set, an audio-visual affect data set, to train and validate the models~\cite{busso2008iemocap}. The dataset comprises of annotations representing both the categorical and dimensional emotional model~\cite{bakker2014pleasure}. To study how the latent representations are transferred between corpora, we use \begin{enumerate*} \item the Surrey Audio-Visual Expresses Emotion~(SAVEE) database that is primarily English and consists of male speakers only, \item the Berlin Database of Emotional Speech~(Emo-DB) recorded in German and, \item the Canadian French Emotional~(CaFE) speech database comprising of French audio samples~\cite{jackson2014surrey, burkhardt2005database, gournay2018canadian}. \end{enumerate*} In this work, we utilize the audio modality only, and constrain the emotional categories to {\it neutral~(N), sad~(S), happy~(H), angry~(A)} as these are the emotions that are common over the data sets. 

\begin{table}[!h]
\centering
\caption{Grouped features of the eGEMAPS~\cite{eyben2015geneva}.}
\resizebox{0.97\columnwidth}{!}{%
\begin{tabular}{ |l|l|l| }
 \hline
 \multicolumn{3}{c}{Features}\\
 \hline

 Pitch & Loudness & Spectral flux (U, UV)\\
 \hline
 Segments/second & Jitter &  Shimmer\\
\hline
  Segment length (UV)& Harmonics & Formant 1~(F1)\\
 \hline
 Formant~2~(F2) & Formant~3~(F3) & Alpha ratio~(V, UV)\\
 \hline
 \pbox{20cm}{Hammerberg\\ Index~(V, UV)} & \pbox{20cm}{Spectral slope 1\\(0-0.5kHz)-(V, UV)} & \pbox{20cm}{Spectral slope 2\\(0.5-1.5kHz)-(V, UV)}\\
 \hline
 \pbox{20cm}{Loudness\\peaks/second} & \pbox{20cm}{Mel-frequency cepstrum\\ Coefficient (MFCC)} & Segment length(UV)\\
 \hline
  \pbox{20cm}{Equivalent\\ sound level} & \pbox{20cm}{Harmonic-to-\\noise-ratio (HNR)}& \\
 \hline

\end{tabular}
}
\label{feature_table}
\end{table}

\noindent
{\bf Features:}
We use the extended Geneva minimalistic acoustic parameter set~(eGeMAPS) in this work~\cite{eyben2015geneva}. Since our objective is to explore the features relevant to the different emotional categories, we used the functionals of lower-level features because summary statistics provide relatively lower variability within a sample. Additionally, while raw speech features may provide higher freedom for the AE to discover representations~\cite{ghosh2015learning, ghosh2016representation}, in such a system it is inherently difficult to understand the association of individual paralinguistic markers to the emotions. Furthermore, emotional annotations in few datasets are provided sample-wise and not segment-wise. Each speech sample yields a feature vector comprising of 88 features and Tab.~\ref{feature_table} illustrates a compressed version of the feature set. We use the OpenSmile toolkit to extract the features~\cite{eyben2010opensmile}. 

\noindent
{\bf Algorithms and Evaluation Metric:}
In this paper, an UAE, that is an AE with the latent dimension much smaller relative to the input feature vector~($z<<x_{input}$), and a DAE are presented for SER~\cite{Goodfellow-et-al-2016}.  
 While AEs have been employed for SER in existing literature, their main focus was to propose network architectures that provide better classification accuracy~\cite{dissanayake2020speech, latif2018variational, neumann2019improving, parthasarathy2019improving, xia2013using}. Since our goal is to explore the cross-corpus transferability of the latent representations and gain insights into the feature markers for the emotional classes, we adhere to the simplest architecture that provides comparable results to previous works.  For a fair comparison between the performance of the UAE and the DAE for SER, we design both systems with identical architectures. The input feature dimension is 88 and we maintain the latent dimension size at $z\in\mathrm{R}^{2\times1}$. In relation to past works, the primary source of low-complexity in our proposed architecture is from the latent dimension size. The network architecture for the UAE is illustrated in Fig.~\ref{fig:archi}, and is similar for the DAE with the only difference being that the input to the DAE is corrupted by a noise component, $x_{\text{input}}=x_{\text{true}}+N$ and $N\in\mathcal{N}(0, 1)$, whereas for the UAE $x_{\text{input}}=x_{\text{true}}$. We use the mean squared error~(MSE) to optimize the network. Since the functionals do not have a temporal correlation and the spatial correlation exists between the statistical parameters of a feature only, we use a fully connected neural network~(NN) instead of a recurrent or convolution type NN.  
 
  \noindent
 For reference, we use principal component analysis~(PCA) to evaluate the quality and the transferability of the latent representations obtained from the UAE and DAE. A linear support vector classifier~(SVC) is used to classify the latent embedding from the PCA, UAE and DAE, and the uncompressed input features, into the four emotional categories. 
We use the unweighted classification accuracy scores to evaluate the performance of the proposed and reference algorithms. The class-wise classification accuracy is demonstrated using the confusion matrix, with the values normalized over the total number of true labels for each emotion. 

\noindent
{\bf Preprocessing:}
Prior to using the data sets for training and testing, we remove the outliers by computing the z-score and eliminating the data samples that have a z-score, $-10>z>10$. We chose a threshold of 10 instead of the standard value of 3 because the goal of this work is to understand the behavior of the models for both typical and atypical rendition of emotions in speech. Therefore, we only remove the extreme outliers. Following this, the data sets are standardized to obtain a normal feature distribution. 

\section{Results and Discussion}
We train and validate the models using 10-fold cross-validation on the IEMOCAP database while the transfer data sets are identical over the iterations. The models were trained over 50 epochs and a batch size of 64, and we used the Adam optimizer with the learning rate set to 1e-3. In the following parts, we evaluate the latent embedding by using them as the input features to classify the speech samples into emotional categories using the SVC. We first present the overall classification accuracy to understand how our presented systems compare to previously proposed methods. Following that, the separability of the emotional categories is investigated and we delve into the physical interpretation of the latent space. At this stage we also investigate how the latent embeddings transfer to unseen corpora. Lastly, we study the input features that pose as the main markers of the emotional categories and demonstrate why certain emotion categories are classified better.

\noindent
{\bf Classification Performance:}
From the classification accuracy presented in Fig.~\ref{fig:acc_emo4}, we observe that for the training and validation data sets, the performance of UAE and DAE is much lower than the accuracy of the reference SVC without any dimensionality reduction. PCA and UAE have similar performances, and the results from the DAE are the lowest on these data sets. The classification accuracy reduces over the transfer data sets, specifically for the German and French data sets. However, the advantage of transfer learning for the UAE and the DAE over the reference methods are visible, as they show the largest classification accuracy in these cases. While the presented architecture in this work is relatively simple~(hidden layers and units, latent dimension, input feature size), we obtain an accuracy that is close to the results presented in existing works~\cite{dissanayake2020speech, latif2018variational, neumann2019improving, parthasarathy2019improving, xia2013using}. 
\begin{figure*}[!tbh]
\centering
\includegraphics[width=0.99\textwidth]{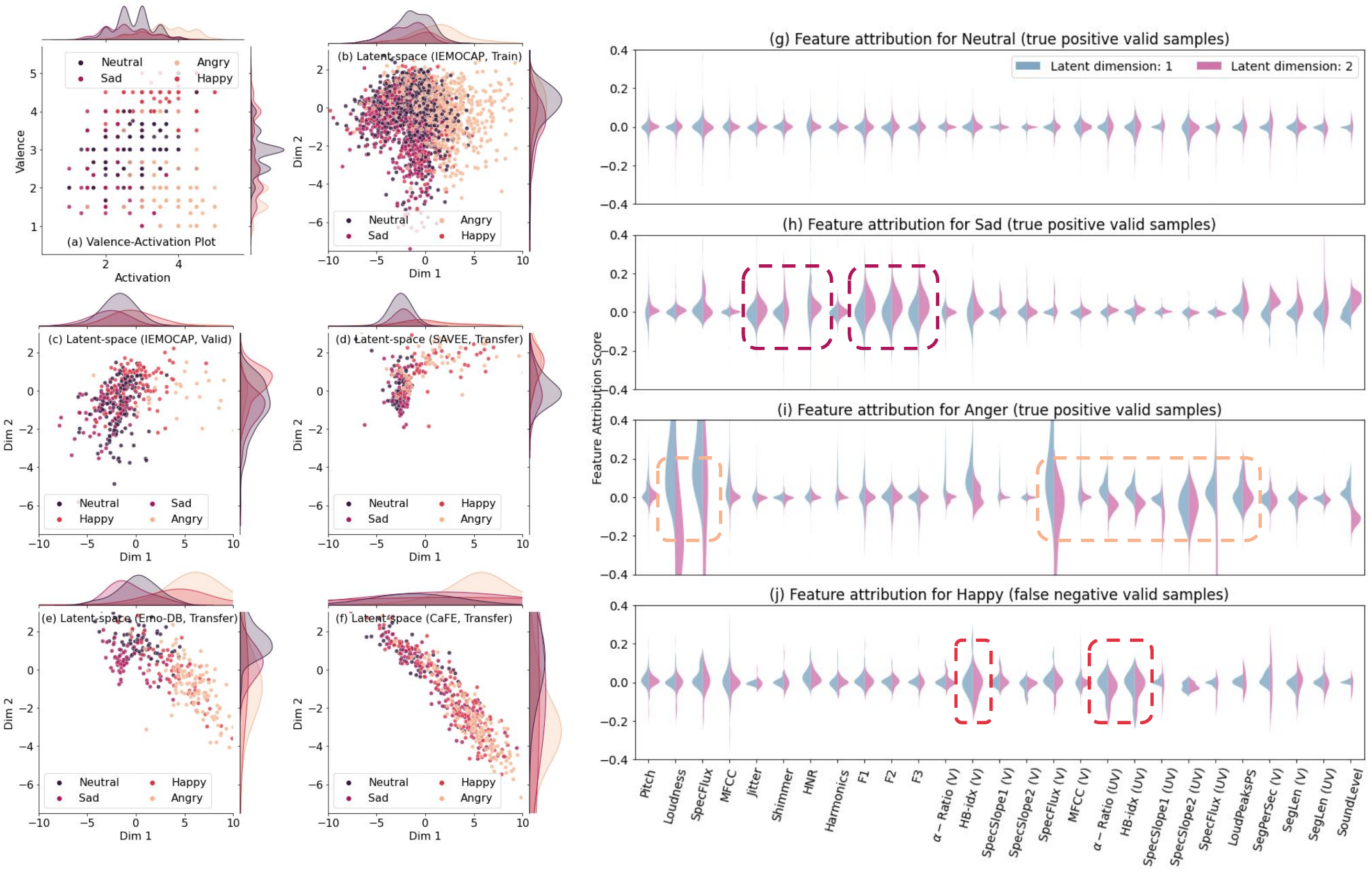}
\caption{(a)-(f)~Scatter plots and marginal distributions of the latent variables from the denoising autoencoder for train, validation and transfer data sets. (g)-(j)~Feature attribution scores using deepLIFT~\cite{shrikumar2017learning} for the latent dimensions over the valid and transfer data sets.}
\label{fig:attr}
\end{figure*}

\noindent
{\bf Analysing Emotions in Triads:}
The results above provide a generic understanding of the performance of the methods under investigation. To gain an insight on the separability of the latent embedding between the four emotions, we process the emotion categories in triads. In other words, we train the UAE and the DAE with samples from three emotional categories at a time. Since an AE operates by minimizing the MSE between the true and reconstructed samples, it is possible that features from an arbitrary emotional category dominates the magnitude of the loss. Therefore, the triad approach also aids in studying if a certain set of features or emotion dominates the training of the AEs. The resulting classification accuracy and the confusion matrices of the DAE, for a randomly selected fold in the cross-validation, are presented in Fig.~\ref{fig:acc_emo3}. We observe that the best results are obtained for the N-S-A triad and the worst results are obtained for the S-H-A triad. Comparing the results over the data sets, results are similar for the English data sets, but the accuracy reduces for German and French. In addition, it can be observed that the emotion {\it happy}~(H) is generally misclassified in all combinations. However, {\it anger}~(A) is consistently classified with high accuracy over all combinations and data sets. Also, unlike for N-S-H, {\it sad}~(S) is classified accurately for S-H-A over the English data sets. 

Following the above observations, we extend the analysis to understand why some emotions are more separable than others and the influence of the differences in corpora on the latent embedding. We do that by visually inspecting the latent representation of the DAE and building a physical interpretation of the latent space by employing the valence and activation labels available in the IEMOCAP data set as the ideal reference for clustering emotions. This is illustrated in Fig.~\ref{fig:attr}~(a) using a scatter plot of the valence and activation values in the training data set, color-coded with the emotion labels. Scatter plots in Fig.~\ref{fig:attr}~(b-f) represent the distribution of the latent embedding for the train, validation and transfer data sets. From Fig.~\ref{fig:attr}~(b-f), we observe that the samples representing {\it anger} form a separate cluster over all the data sets, whereas samples representing {\it neutral} and {\it sad} overlap and form a joint cluster in the latent space. In contrast the {\it happy} samples are spread over both of these clusters. Furthermore, the marginal distribution of the classes for the activation variable in Fig.~\ref{fig:attr}~(a) resembles the marginal distribution of Dim 1 shown in Fig.~\ref{fig:attr}~(b-f). We therefore conclude that the DAE learns to cluster the samples in terms of their activation. Additionally, while the latent clusters of the training, validation and transfer data sets are similar, the orientation of the distributions are rotated for the German and French data sets relative to the English data sets. This explains the observed reduction in the classification accuracy over the German and French data sets in Figs.~\ref{fig:acc_emo4}~and~\ref{fig:acc_emo3}.

\noindent
{\bf Feature Attribution for Emotion Clustering:}
In the final part of the analysis, we study the features that pose as the main markers for each emotional category. To that end, we utilize the DeepLIFT~(Deep Learning Important FeaTures) algorithm that computes the influence of the input features on a specific neuron through the difference in the output, relative to the reference output when there is a difference in input between the reference and the considered input sample~\cite{shrikumar2017learning}. We computed the mean input feature vector over the true positives from the neutral samples of the validation set and employed that as the reference. The resulting distributions of feature attribution scores for the latent dimensions are demonstrated as violin-plots in Fig.~\ref{fig:attr}~(g-j) over the true positives from the validation set. Note that we have grouped similar features for effective visualization. 

The attribution scores of the {\it neutral} class is similar over all features, as shown in Fig.~\ref{fig:attr}~(g). This is expected since our reference is the neutral class. The classification accuracy and the confusion matrices have indicated that the clustering of {\it anger} is consistently accurate. This is also reflected in Fig.~\ref{fig:attr}~(i) where we observe that the features with high attribution scores are unique to this emotion class, specifically features related to loudness, spectral flux and spectral slope. Although we observe that the formants~(F1, F2, F3), jitter, shimmer and harmonic-to-noise-ratio~(HNR) have a relatively high feature attribution and are unique to {\it sad}, classes {\it sad} and {\it happy} seem to generally share features with high attribution scores as shown in Fig.~\ref{fig:attr}(h,~j). Furthermore, the observation from the classification accuracy indicates that samples representative of the emotion class {\it happy} are particularly difficult to cluster and classify. This can be explained from the feature attribution scores in Fig~\ref{fig:attr}(j) that are very similar to the scores of the neutral class. In other words, the proposed DAE architecture is unable to distinguish between features from {\it neutral} and {\it happy} classes. From this analysis we find the specific features that are directing the SER decisions in the proposed DAE and this understanding can be employed when the model is used on newer and unseen corpora, for instance on a cross-cultural emotion data set, to interpret the model and introduce transparency into the decision process.  

\begin{comment}
Methods for feature selection:
\begin{enumerate}
\item Laplacian score
\item Multi-cluster Feature Selection
\item Unsupervised discriminative feature selection
\item Autoencoder feature selection
\item Integrated gradients, layer-wise back (relevance) propagation, occlusion (occlusion sensitivity and maps), deep taylor, captum
\end{enumerate}
\end{comment}

\section{Conclusion}

In this work, we have proposed a low-complexity UAE and DAE that achieves accuracy rates similar to existing methods, at a relatively lower computation cost due to a highly compressed latent space. The goal of this work was to make the system interpretable and study the transferabilty of the latent representation, whereby we performed an in-depth analysis of the latent representations and their physical interpretation. We learned that the model implicitly learns to cluster the emotion classes according to their activation levels. Additionally, we observe that the orientation of the latent distributions for the German and French data sets is different to that of the English data sets. Following this, we conducted a study on the main feature markers for each emotion class. We discovered that the highest feature attribution scores are obtained for the class {\it anger} and are unique for this class, whereas the feature attributions for {\it happy} are very similar to the reference neutral category.

\bibliographystyle{IEEEtran}
\balance
\bibliography{ref}
%\nocite{*}
\end{document}